\documentclass[aps,pra,nopacs,preprint]{revtex4-2}
\usepackage{graphicx}
\usepackage{subfig}
\usepackage{tabularx}
\usepackage{bm}
\usepackage{amsmath}
\usepackage{esvect}
\usepackage{lineno}

\providecommand{\abs}[1]{\left\lvert#1\right\rvert}

\providecommand{\bra}[1]{\langle #1 \rvert}
\providecommand{\ket}[1]{\lvert #1 \rangle}

\providecommand{\be}{\begin{equation}}
\providecommand{\ee}{\end{equation}}
\providecommand{\ba}{\begin{eqnarray}}
\providecommand{\ea}{\end{eqnarray}}

\usepackage{mathbbol}

\usepackage{tikz}
\usepackage{amsfonts,amssymb}
\usepackage{dsfont}
\usepackage{physics}
\usepackage{tocvsec2}
\usepackage{amsthm}

\usepackage{graphicx}
\usepackage{subfig}
\usepackage{tabularx}
\usepackage{bm}
\usepackage{amsmath}
\usepackage{esvect}

\usepackage{siunitx}
\usepackage{hyperref}
\usepackage[capitalize]{cleveref}

\providecommand{\abs}[1]{\left\lvert#1\right\rvert}

\providecommand{\bra}[1]{\langle #1 \rvert}
\providecommand{\ket}[1]{\lvert #1 \rangle}

\providecommand{\be}{\begin{equation}}
\providecommand{\ee}{\end{equation}}
\providecommand{\ba}{\begin{eqnarray}}
\providecommand{\ea}{\end{eqnarray}}
\usepackage{amsfonts,amssymb}
\usepackage{dsfont}
\usepackage{physics}
\usepackage{tocvsec2}

\hypersetup{colorlinks=true,linkcolor=blue,citecolor = blue, urlcolor=black}
\newcommand{\beq}{\begin{equation}}
\newcommand{\eeq}{\end{equation}}

\usepackage{appendix}

\begin{document}

\title{Phase-space representation of classical simulability for arbitrary bosonic encodings}

\author{Eloi Descamps$^{1}$}
\author{Astghik Saharyan$^{1}$}
\author{ Arne Keller$^{1,2}$}
\author{Pérola Milman$^{1}$ }
\email{corresponding author: perola.milman@u-paris.fr}

\affiliation{$^{1}$Université Paris Cité, CNRS, Laboratoire Matériaux et Phénomènes Quantiques, 75013 Paris, France}
\affiliation{$^{2}$Département de Physique, Université Paris-Saclay, 91405 Orsay Cedex, France}

\begin{abstract}

Phase-space representations can be used to reveal resources for quantum computational advantage. In odd-dimensional qudit systems, states and operations admitting non-negative quasi-probability representations are efficiently classically simulable, and analogous conclusions hold for bosonic systems, usually represented in quadrature phase space. However, when quantum information is encoded in bosonic systems, the link between physical phase-space representations and computational resources becomes nontrivial: large negativities may arise in the physical space even when the encoded model remains efficiently simulable. We establish a general framework connecting physical phase spaces of bosonic systems to the effective computational phase spaces associated with arbitrary encodings. We show that bosonic Hilbert spaces with finitely many modes can always be exactly encoded on a quantum computer, without truncation or approximations, and that for any encoding there exists an adapted physical phase-space representation in which all provably classically simulable states and protocols are non-negative. Conversely, negativity can always persist in physical representations for arbitrary encodings. Our results formalize the operational meaning of phase-space negativity in bosonic quantum information and its relation with computational resources.
\end{abstract}

\pacs{}
\vskip2pc 
\maketitle
\section{Introduction}

Bosonic systems are one of the most promising platforms for quantum information protocols. Several physical platforms now provide well-controlled and environmentally isolated set-ups supplemented with error correcting codes both in the discrete- and continuous variable regimes \cite{eczoo_oscillators, PhysRevX.6.031006, PhysRevA.56.1114, gottesman_encoding_2001, PhysRevLett.133.240603, albert_2025}. Another key example is boson sampling \cite{BosonSampling}, one of the few quantum computational tasks strongly believed to exhibit quantum advantage, defined by the impossibility of efficient classical simulation. Identifying practical and general methods to represent, quantify, and certify bosonic resources that can—or provably cannot—enable such advantage is therefore of central importance \cite{calcluth2023sufficient, PhysRevA.104.012430, ModularMenicucci, PhysRevLett.130.090602, descamps2024superselectionrules, descamps2025unifiedframeworkbosonicquantum}.

A particularly powerful approach to identifying resources for quantum computational advantage relies on phase-space methods \cite{PhysRevLett.117.180401, PhysRevA.59.971, Hillery1984DistributionFunctions}, which provide an alternative representation of quantum states via the Stratonovich–Weyl correspondence. Within this framework, necessary conditions for quantum advantage can be expressed in terms of properties of the Wigner function \cite{PhysRevLett.109.230503, VeitchEtAl2012, Veitch2014, Bartlett2002, Bartlett2002Efficient}, in close analogy with the Gottesman–Knill theorem  \cite{gottesman1998heisenberg}. In particular, negativity of the Wigner function is identified as a necessary resource for quantum computational advantage.

In continuous-variable systems, described in quadrature phase space, such negativities are closely linked to non-Gaussian features of states and operations through Hudson’s theorem \cite{HUDSON1974249, Soto:1983avg}. Similar results hold for discrete systems \cite{Ernesto, Pashayan2015, Gibbons2004}, where a discrete version of Hudson’s theorem applies to odd-dimensional qudits \cite{Gross2006Hudsons, Gross}, and stabilizer states appear as non-negative distributions.

However, while qudits describe distinguishable systems, bosonic platforms involve indistinguishable particles and require encoding quantum information into physical states and degrees of freedom. As a result, the connection between physical phase-space representations and computational resources becomes nontrivial. In particular, states and operations that are efficiently classically simulable in the encoded (computational) spaces, and appear as nonnegative distributions in the computational phase space, may nevertheless exhibit strong negativities in the quadrature (physical) phase space  \cite{xmtw-g54f, Hahn2025classicalsimulation,Calcluth2022efficientsimulation}.

This observation has motivated several works \cite{PhysRevA.102.022411, davis2024, PRXQuantum.6.010330, PhysRevA.104.022408} aiming to identify bosonic resources for quantum computation using adapted phase-space representations. Ideally, such representations would ensure that all efficiently simulable computational processes correspond to non-negative distributions in the physical phase space. Nevertheless, this has not yet been achieved, and existing approaches face two main limitations. First, bosonic modes are associated with infinite-dimensional Hilbert spaces, whereas qudit encodings are finite-dimensional, preventing a bijective mapping between physical and computational spaces  \cite{Andreas, PRXQuantum.6.010330, maltesson2025}. Second, existing constructions are typically encoding-dependent—often motivated by quantum error correction—and do not provide a unified and general connection between computational simulability and physical phase-space representations for arbitrary encodings.

In this work, we address the two limitations identified above using a general framework that systematically connects phase-space representations of physical bosonic systems with the effective computational phase spaces induced by arbitrary encodings. First, we construct an explicit bijective mapping between the full bosonic Hilbert space of a finite number of modes and a quantum computational space, enabling an exact representation of bosonic states and operations without truncation. This shows that finite-mode bosonic systems can be treated as exact computational spaces, rather than approximate ones, with no inconsistencies. Second, for any given encoding, we identify a corresponding adapted bosonic phase-space representation in which all states and operations that are efficiently classically simulable are represented by non-negative quasi-probability distributions. This establishes an operational connection between classical simulability and phase-space nonnegativity directly at the level of the physical system. 

A central ingredient of our construction is the explicit identification of a quantum phase reference, defined consistently for both the encoded (computational) space and the underlying bosonic system. We show that in the standard quadrature phase-space representation, this reference is provided by the vacuum state. More generally, however, arbitrary states can be chosen as phase references, leading to generalized planar phase-space representations beyond the conventional quadrature framework and naturally accommodating different encodings.

\section{Results}
\subsection{Superselection rule compliant representation and the continuous variables limit}\label{sec:SSRC}

The usual approach to represent bosonic states consists as considering them as superpositions of Fock states. In this framework, the general form of the simplest state, defined on a single mode labelled $a$, is given by 
\be\label{CVstate}
\ket{\psi}_c = \sum_{n=0}^{\infty} c_n \ket{n}_a, 
\ee
with $\sum_{n=0}^{\infty}\abs{c_n}^2=1$ and $\ket{n}_a=\frac{\left (\hat a^{\dagger} \right )^n}{\sqrt{n!}}\ket{\emptyset}$, where $\ket{\emptyset}$ denotes the vacuum $\hat a\ket{\emptyset}=0$ . As discussed in the literature, such states formally violate the particle-number superselection rule (SSR) \cite{PhysRevA.68.042329, Sanders_2012, RevModPhys.79.555, doi:10.1142/S0219749906001591, PhysRevA.55.3195}, implicitly assuming the existence of an external phase reference that defines coherence between different Fock states. Instead, one can adopt a more general representation that is SSR compliant (SSRC), and explicitly includes the phase reference as a quantum resource, enforcing total particle-number conservation. The general minimal form of a pure SSRC state is
\begin{equation}\label{state}
\ket{\psi}_s = \sum_{n=0}^N c_n \ket{n}_a \ket{N-n}_b, 
\end{equation}
where modes $a$ and $b$ serve as mutual phase references. States in the form \eqref{CVstate}, termed from now on the continuous variables (CV) representation, can be recovered as a limit of the SSRC representation when coefficients $c_n$ in \eqref{state} are such that there is a large photon-number imbalance between modes $a$ and $b$, such that ${}_s\bra{\psi}\hat{a}^\dagger \hat{a}\ket{\psi}_s \ll \sqrt{N}$, with $\hat{a}^\dagger \ket{n}_a = \sqrt{n+1}\ket{n+1}_a$ \cite{descamps2025unifiedframeworkbosonicquantum, moulonguet2026}. The identification of this imbalance - hence, the subspace associated with the CV representation - is always relative to a chosen modal basis—in Eq.~\eqref{state}, modes $a$ and $b$. In this sense, the vacuum in the CV formalism is rather a phase reference than the absence of photons, and it corresponds to state $\ket{N}_b$

\begin{figure}[h]  
    \centering
    \includegraphics[width=\linewidth]{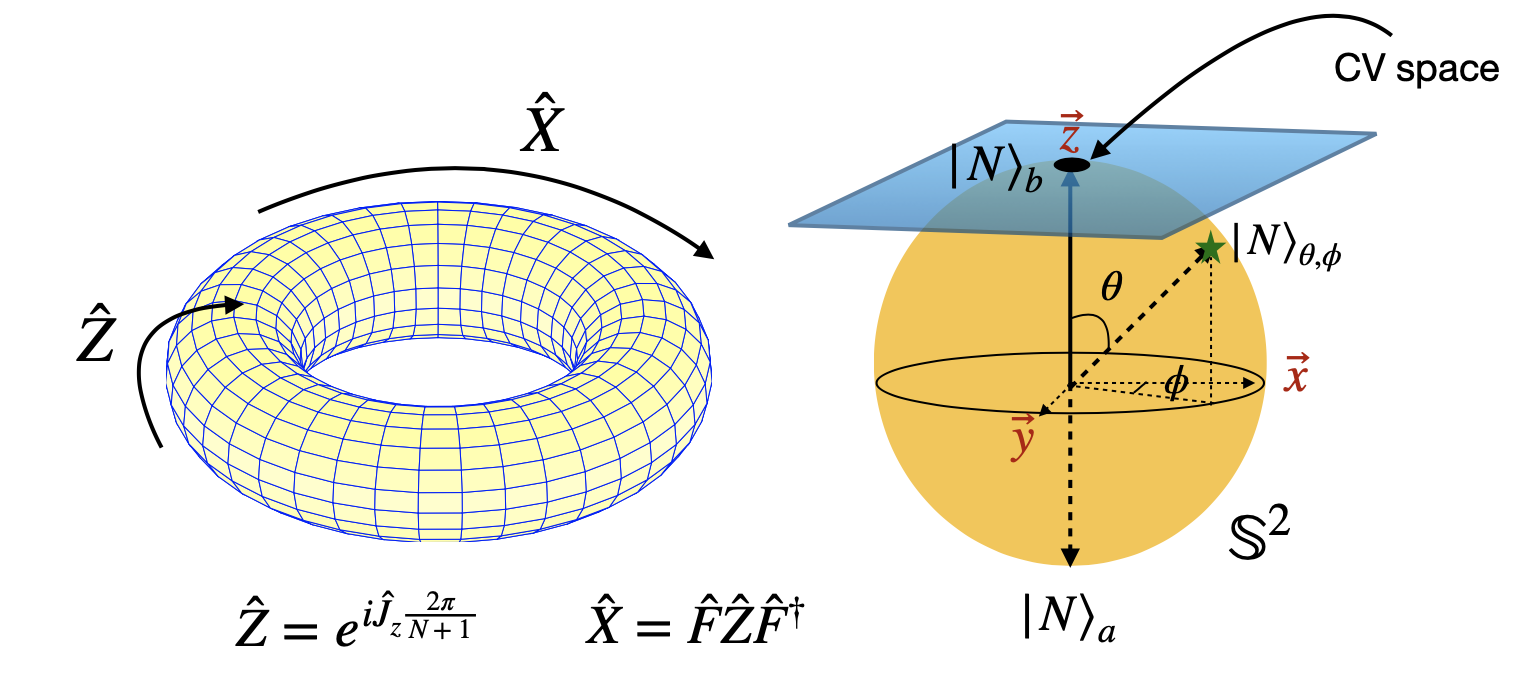}  
    \caption{
        (Color online) Intuitive illustration of the three phase space representations considered in this work. Right: SSRC bosonic states admit a spherical representation, such as the spherical Wigner function, which depends on the choice of a orientation (the physical phase reference). Coherent states, Fock states, and more generally particle-separable states fix a direction on the sphere, with the extremal eigenstates of $\hat J_{\vec n}$ occupying antipodal points along $\vec n$. The tangent plane at a chosen point on the sphere corresponds to the continuous variable (CV) limit ($\theta \ll 1$), and the point of tangency defines the vacuum of the planar CV phase space, given by its vicinity. Left: A qudit discrete phase space is constructed from generalized Pauli operators built from SSRC bosonic unitaries, providing the computational phase space description used throughout the work. }
    \label{fig1}
\end{figure}

The limit relating \eqref{state} to \eqref{CVstate} can also be geometrically represented, as shown in Fig. \ref{fig1}: The states $\ket{n}_a \ket{N-n}_b$ can be mapped into angular-momentum eigenstates $\ket{j,m}$ with $j = N/2$ and $m = n - N/2$, or equivalently into symmetric states of $N$ two-level bosonic particles, and the operational connection between angular-momentum and bosonic systems is established by the Schwinger representation: $\hat{J}_z = \tfrac{1}{2}(\hat{b}^\dagger \hat{b} - \hat{a}^\dagger \hat{a})$, $\hat{J}_x = \tfrac{1}{2}(\hat{a}^\dagger \hat{b} + \hat{b}^\dagger \hat{a})$, $\hat{J}_y = \tfrac{i}{2}(\hat{a}^\dagger \hat{b} - \hat{b}^\dagger \hat{a})$ \cite{Schwinger}. Hence, states in the SSRC representation can be represented by a spherical Wigner function \cite{PhysRevResearch.3.033134,KOWALSKI2023169428, PhysRevA.49.4101, PhysRevA.63.012102, app15095155, Klimov_2017} (see Eq.~\eqref{WSphere} in Methods and right panel of Fig.~\ref{fig1}). Different Fock states (particle-separable states, as spin-coherent states \cite{JMRadcliffe_1971}) appear as distributions centered at different points on the sphere, corresponding to different spatial directions (or modal basis) as indicated in Fig.~\ref{fig1}. The two $\hat J_z$ eigenstates with maximal and minimal eigenvalues are located at antipodal points and rotations around directions $\vec n$, $e^{-i \hat J_{\vec n}\zeta}=e^{-i\hat J_z\varphi}e^{-i\hat J_y\theta}\equiv \hat R(\theta,\varphi)$, transform them as $ \hat R(\theta,\varphi) \ket{N}_b= \ket{N}_{\theta,\varphi} = \sum_{n=0}^N \sqrt{\binom{N}{n}} \left(\cos\frac{\theta}{2}\right)^{N-n} \left(e^{i\varphi}\sin\frac{\theta}{2}\right)^n \ket{n}_a \ket{N-n}_b$,  which remains particle-separable. Rotations are passive transformations: for any state of the
form Eq. \eqref{state}, they merely change the mode basis and therefore do not alter the intrinsic particle properties of the state. Within this geometric picture, the CV limit corresponds to a  restriction to the tangent plane of the sphere ({\it i.e.}, the region
$\theta \ll 1$), as illustrated in Fig.~\ref{fig1} and discussed in \cite{RevModPhys.90.035005}. In this limit, the spherical Wigner function reduces to the familiar planar Wigner function in quadrature phase space
\cite{PhysRevA.63.012102} (see Eq.~\eqref{WPlanar} in the End Matter), under the correspondence $\hat J_x \rightarrow \sqrt{N} \hat x$, $\hat J_y \rightarrow \sqrt{N} \hat p$, $\hat J_z \rightarrow \frac{N}{2} \mathbb{1}$, with $\hat x=\frac{1}{2}\left ( \hat a+\hat a^{\dagger}\right )$ and $\hat p=\frac{1}{2i}\left ( \hat a-\hat a^{\dagger}\right )$. Consequently, the quadrature eigenstates $\ket{x}$ and $\ket{p}$ represent the CV limit of the eigenstates of $\hat J_x$ and $\hat J_y$, respectively. Also, in the CV limit, SSRC infinitesimal rotations generated by $\hat J_{\vec{n}}$, $\vec{n}\perp \vec{z}$, act as displacements in the planar quadrature phase space \cite{descamps2025unifiedframeworkbosonicquantum}, as we recall in the Supplementary Material for completeness \cite{SM}. In this case, the mode $b$'s population is effectively considered as $N$ in calculations (since $\hat b^{\dagger}\ket{N-n}_b \approx \sqrt{N}\ket{N-n+1}_b$), even though the representation Eq. \eqref{state} still holds. 
As mentioned, the CV limit implicitly selects a phase reference, or equivalently, a mode basis, or quantization axis. This is represented in the planar phase space as a choice of vacuum, defining the origin of phase space: for $\theta \ll 1$ (Fig. \ref{fig1}), the state $\ket{0}_a \ket{N}_b$ is associated with the single-mode vacuum $\ket{0}_a$ in the quadrature representation. It is important to notice that once the vacuum is chosen, states that can be represented in the associated planar phase space correspond to the vicinity of the point on $\mathbb{S^2}$ that defines the tangent plane. Within this framework, a natural generalization of the notion of vacuum arises by selecting a direction $(\theta,\varphi)$ and defining the corresponding vacuum as the rotated state $\hat R(\theta,\varphi)\ket{N}_b \equiv \ket{N}_{\theta,\varphi}$. In this representation, CV states are identified with those supported on the plane tangent to this reference point, namely states satisfying ${}_s\langle \psi | \hat R(\theta,\varphi)\,\hat a^\dagger \hat a\,\hat R^\dagger(\theta,\varphi) | \psi \rangle_s \ll \sqrt{N}$. These states define a subspace distinct from the conventional CV sector associated with the standard vacuum $\ket{0}_a\ket{N}_b$. A key consequence of our approach is that even this simple construction—where reference-frame transformations are implemented solely by rotations and any state $\hat R(\theta,\varphi)\ket{N}_b$ may serve as a vacuum—already extends beyond the usual CV framework. While different planar phase spaces can be defined through alternative Wigner kernels \cite{PhysRevA.59.971}, such constructions are typically formulated in an infinite-dimensional Hilbert space $L^2(\mathbb{R})$ \cite{PhysRevA.102.022411,PhysRevResearch.5.L032006} and do not incorporate the phase-reference mode. By contrast, the transformations considered here act on the full finite-dimensional Hilbert space $\mathcal{H}_{N+1}$, within which CV behavior emerges only as a limit. This motivates our use of a SSRC representation, which remains fully finite-dimensional and more general in terms of accessible transformations, while recovering the standard CV description in the appropriate regime.

Finally, we highlight an additional structural feature of the CV limit. As shown in \cite{moulonguet2026}, physical states in this regime—namely, those corresponding to phase-space distributions normalizable in the vicinity of the point defining the tangent plane—are precisely those that can be approximated arbitrarily well within a Hilbert space of dimension $K \ll \sqrt{N}$. This shows that the CV limit does not correspond to an enlargement of the bosonic Hilbert space, but rather to a restriction to a low-dimensional subspace. More precisely, the CV regime is defined by a constrained limit in which $K \to \infty$ and $N \to \infty$, while maintaining $K \ll \sqrt{N}$. 

\subsection{Information encoding}\label{sec:Info}

We now introduce a systematic encoding framework that will enable us, in Section~\ref{sec:Phase}, to address the two main challenges identified above: relating physical and computational resources, and providing a physical phase-space representation of this connection. Specifically, in this Section, we will:
\begin{enumerate}
\renewcommand{\labelenumi}{\alph{enumi})}
\renewcommand{\theenumi}{\alph{enumi}}
\item \label{A} Show that the full bosonic Hilbert space (including its CV representations) with a finite number of modes can be arbitrarily well approximated by a quantum computer, without truncation, and vice-versa, a quantum computer can always be used to efficiently simulate the full bosonic space with a finite number of modes.
\item \label{B} Identify, for each encoding of a quantum computer using bosonic systems, what are the bosonic resources (as for instance quadrature- or SSRC Gaussian and non-Gaussian features) required to implement elements of the quantum computer, including the computational basis and a universal gate set.
\end{enumerate}

For this we start by recalling that 
\be\label{Uni}
\mathcal{U}_S=\{ e^{i \hat J_{\vec{n}} \theta},\, e^{i \hat J_{\vec{n}}^{2} \chi} \}
\ee
is a universal set in the SSRC representation \cite{PhysRevLett.132.153601,descamps2024superselectionrules,descamps2025unifiedframeworkbosonicquantum}, where $e^{i \hat J_{\vec{n}}^{2} \chi}$ is analogous to a spin-squeezing operator. The set $\mathcal{U}_S$ generalizes the corresponding CV universal set $\mathcal{U}_C=\{ e^{i \hat x q},\, e^{i \hat x^{2} q},\, e^{i \frac{\pi}{4} (\hat x^{2} + \hat p^{2})},\, e^{i\hat x_i\otimes \hat x_j}, e^{i \hat x^{3} q} \}$ \cite{Braunstein}, which is recovered in the CV limit. The unitary  $e^{i \hat J_{\vec{n}}^{2} \chi}$ is a non-Gaussian operation in both the quadrature and SSRC representations, while rotations $e^{i \hat J_{\vec{n}} \theta}$ are Gaussian in both frameworks, and correspond to a mode basis change, as previously discussed. Consequently, $\mathcal{U}_S$ naturally partitions in two types of operations that we define as SSRC Gaussian (SG) and SSRC non-Gaussian (SNG) operations. The advantage of working with $\mathcal{U}_S$ is that it clearly identifies all non-rotation operations as SNG, enabling a transparent classification of the physical resources required for information encoding and computation, permitting showing point \ref{B} above.

In \cite{descamps2025unifiedframeworkbosonicquantum}, such a partition was used to identify physical resources associated with encoding qubits into bosonic states, thereby enabling the representation of a two-dimensional subspace of the full bosonic Hilbert space as a quantum computational space. Here, we address a different problem: rather than restricting to a subspace, we construct a mapping of the full bosonic Hilbert space onto a quantum computer. Within the SSRC representation adopted in this work, the bosonic Hilbert space has finite dimension $N+1$—which can be made arbitrarily large—allowing for an exact mapping onto a qudit of dimension $d = N+1$, (we will always consider $d$ as odd unless specified), without requiring any constraints or continuous-to-discrete mappings \cite{maltesson2025, PRXQuantum.5.010331, chabaud2025energybosonscomputationalcomplexity, xmtw-g54f, orlov2025}. To characterize the possible mappings, we build on the formal framework developed in \cite{AVourdas_2004}, previously employed in \cite{PhysRevA.65.052316} to study specific qudit encodings on finite-dimensional subspaces arising in the continuous-variable limit. After introducing our general construction, we compare it with these earlier approaches, highlighting subtle but important conceptual differences.

Pauli-like operators $\hat X$ and $\hat Z$ can be defined on a Hilbert space of dimensions $d$ satisfying the Heisenberg-Weyl commutation relation $\hat X^l \hat Z^m = \omega_d^{-lm}\, \hat Z^m \hat X^l$ for $l,m\in\mathbb{Z}_d$, where $\omega_d = e^{i 2\pi/d}$. These operators act on the qudit computational basis $\{\ket{\overline j}\}$ as $\hat X \ket{\overline j} = \ket{\overline {(j+1) \bmod d}}$ and $\hat Z \ket{\overline j} = \omega_d^{j}\ket{\overline j}$. From these definitions, qudit universal gate sets can be constructed~\cite{PhysRevA.71.052318,PhysRevA.95.062303}, and a discrete Wigner function on a discrete toric phase space can be associated with the code~\cite{PhysRevA.65.062309,tilmawigner2016,ALuis_1998,BIANUCCI2002353,PhysRevA.15.449,Gross,s5wnmysr,Chaturvedi2006,Marchiolli_2019} (see Fig. \ref{fig1} and  Eq.~\eqref{WQudit} in Methods). 

Following~\cite{AVourdas_2004}, one may for instance identify $\hat Z = e^{i \hat J_z 2\pi/(N+1)}$ and $\hat X = \hat F^{\dagger} \hat Z \hat F = e^{i \hat \theta_z 2\pi/(N+1)}$, where $\hat F = \sum_{n,m=0}^{N} \bigl( \omega_{N+1}^{nm}/\sqrt{(N+1)} \bigr)\ket{n}_a\ket{N-n}_b\, {}_b\!\bra{N-m}\, {}_a\!\bra{m}$, and where $e^{i\hat\theta_z 2\pi/(N+1)}$ is the relative-phase operator~\cite{PhysRevA.48.4702} and the computational basis is encoded as $\ket{j}_a\ket{N-j}_b \mapsto \ket{\overline j}$. Importantly, for this encoding, and using $\mathcal{U}_S$ \eqref{Uni} we conclude that  $\hat X$ is necessarily an SNG operation since it is not a rotation. Interestingly, since the mapping between the physical basis and the qudit basis is one-to-one, the universal gate set in the qudit space is also universal in the bosonic SSRC space, allowing one to approach arbitrarily well any state of the form~of Eq. \eqref{state} using the qudit's universal gates. We have then showed two results: it is always possible to exactly map the bosonic space into a quantum computer, on one hand, and on the other hand, a quantum computer can always be used to efficiently simulate the full bosonic space with a finite number of modes, showing point \ref{A} above. Consequently, any state in the bosonic space admits a corresponding Wigner function in the qudit's discrete toric phase space. Finally, this code construction applies to any orthogonal basis of the $(N + 1)$-dimensional, two mode, $N$-bosons Hilbert space - including for arbitrarily large $N$ and in the CV limit. 

More specifically, these two last results are independent of the encoding: this means that one can perfectly well choose different encodings 
\be\label{U}
\hat U\ket{j}_a\ket{N-j}_b \mapsto \ket{\overline j},
\ee 
where $\hat U$ is an arbitrary unitary SSRC operation, and construct the associated universal gate set and Pauli-like operators accordingly. Importantly, for being an arbitrary SSRC operation, the possible $\hat U$ include unitary operations that cannot be defined in the CV limit for manipulating the phase reference mode in a non-trivial way - the rotation example discussed in Section \ref{sec:SSRC} being one example, and others will be discussed in Section \ref{sec:Phase}.

The possibility of exact mapping of the full multimode bosonic space into a quantum computer that efficiently simulates it is an important tool that will be later used to derive the main results of this paper. Yet, it is worth discussing it and comparing it with previous results so as to exhibit the new aspects of considering the SSRC representation. In \cite{PhysRevA.65.052316}, for instance, two encoding procedures have been proposed, also from \cite{AVourdas_2004}, and both present similar characteristics with respect to the issues we now discuss. For this reason, we first limit the discussion to the first encoding proposal. In this encoding, Fock states (say, in mode $a$) are mapped into qudits, $\ket{j}_a \mapsto \ket{\overline j}$ and the corresponding Pauli-like operators are introduced. They also act on a space of finite dimension, $d$, which is now a subspace of the infinite dimensional Hilbert space ${\cal H}_{\infty}$. In such an encoding, we have that $\hat X_a = \sum_{n=0}^{d-1} \ket{n+1}_a{}_a\bra{n}+\ket{0}_a{}_a\bra{d}$, which is the Pegg-Barnett phase operator \cite{DTPegg_1988}. A first important difference between \cite{PhysRevA.65.052316} and the present approach is that states $\ket{j}_a$ and operators $\hat X_a$ used in \cite{PhysRevA.65.052316} are defined in the single-mode Fock space, which corresponds to the CV limit of the SSRC representation, as discussed. For this reason, the authors comment on the difficulties of the $d\to \infty$ limit, that is considered as necessary so as to encode the full single mode bosonic space into a quantum computer. In particular, they remark that $\hat X_a$ becomes problematic in this limit, which is indeed the case if one restricts to the CV representation. Notice that this is not the case in the SSRC representation, where the logic gates ($\hat X$ and $\hat Z$, for instance) are always well defined, for any $N$. Interestingly, we can notice that Pegg and Barnett  \cite{DTPegg_1988} propose to solve the problem of the definition of $\hat X_a$ when $d \to \infty$ by considering that the limit is taken only after the computation of relevant $c$-number physical quantities, which corresponds to the result of physical measurements involving this operator. This approach is similar to the procedure adopted here to determine the subspace corresponding to the CV limit, but does not include the restrictions associated with the CV limit as we explicit here. In any case, the encodings proposed in \cite{PhysRevA.65.052316} do not enable mapping the full single mode bosonic space into a quantum computer, which is, on the contrary, possible using the SSRC representation. As a matter of fact, when considering single-mode states in the CV representation, the encodings proposed in \cite{PhysRevA.65.052316} already exhibit structural and dimensional limitations compared to the SSRC representation, preventing the encoding of the full bosonic Hilbert space. The underlying reason for this is that the CV limit, as analyzed in detail in \cite{moulonguet2026}, involves a richer physical and mathematical structure than the naive limit $d \to \infty$. In particular, as discussed in Section \ref{sec:SSRC}, this limit is constrained: one must also impose $d \ll \sqrt{N} \to \infty$ (with $d = K$ in the notation of Section \ref{sec:SSRC}). These constraints are essential for a consistent definition of states in ${\cal H}_{\infty} = L^2(\mathbb{R})$ and have direct implications for information encoding and the associated bosonic resources. Consequently, the notation $\ket{j}_a$ cannot be regarded as a simple relabeling of $\ket{j}_a\ket{N-j}_b$. We will return to these points later, where we extend and generalize CV encodings while clarifying the role of the physical resources required for their implementation.

\subsection{Phase space representations and bosonic resources for quantum information}\label{sec:Phase}

We now arrive at the central result of this work. By combining the tools introduced above, we establish the following:

\begin{enumerate}
\item \label{one} For any encoding of the bosonic Hilbert space into a quantum computer, there exists, in the continuous-variable limit, a planar (quadrature-like) quasi-probability representation of phase space in which all efficiently classically simulable architectures are described by non-negative distributions.
\item \label{two} Conversely, for any such encoding, there also exist planar phase-space representations in which these same efficiently simulable protocols exhibit negativities.
\end{enumerate}

To prove these statements, we consider the encoding \eqref{U}. This encoding induces a map from the bosonic spherical continuous phase space to the computational qudit toric discrete phase space and consequently a mapping between the corresponding Wigner functions. Crucially, the resulting computational discrete phase-space representation is independent of the specific physical resources used to define the computational basis and logical operations. Although $\hat U$ may correspond to distinct classes of physical transformations (SG or SNG), thereby associating the computational basis with different physical resources, it does not modify the phase-space structure of the encoded qudit. Instead, properties of the discrete phase space—such as the presence or absence of negativity in the discrete Wigner function—are determined solely by the states and operations defined in the code space, independently of the involved physical resources. Consequently, physical operations that are SNG or SG may be mapped to either Clifford or non-Clifford gates in the computational representation. The emergence of negativity in the associated toric phase-space distributions \cite{AVourdas_2004, WOOTTERS19871, Albert_2017} therefore depends on the code, rather than on the physical nature of the underlying bosonic operations.

We now address the representation of the physical resources associated with the code on a planar phase space, corresponding to the CV limit. As discussed, the CV limit and the associated Wigner function appear from the spherical SSRC representation as a a tangent plane to a point that defines the vacuum, or the phase reference associated with the planar representation.

We start by introducing an arbitrary SSRC unitary operator $\hat K \in SU(N+1)$, and observe that the universal set ${\cal U}_S$ can be replaced by any unitarily equivalent one, ${\cal U}_{KS}=\{ e^{i\hat{\mathcal J}_{\vec n}\theta},\,e^{i\hat{\mathcal J}_{\vec n}^{\,2}\chi}\}$, where $\hat{\mathcal J}_{\vec n}=\hat K\,\hat J_{\vec n}\,\hat K^{\dagger}$. In general, we will consider $\hat K$ as a unitary operation that is associated with transformations of the physical space, {\it i.e.}, the basis used to represent the spherical Wigner function \eqref{WSphere}. It can then be treated independently of $\hat U$ that determines the mapping of a given physical basis (set by a choice of $\hat K$) into the qudit space. 
In the unitarily transformed representation, we can define $\hat K\ket{n}_a\ket{N-n}_b=\ket{n, N-n}_K$, so that $\hat{\mathcal J}_{z}\,\hat K\ket{n}_a\ket{N-n}_b=(N-2n)/2\ket{n, N-n}_K$ (the notation is further explained in Methods). This new basis spans the same $(N+1)$-dimensional bosonic space as the original one ($\hat K=\mathbb{1}$), and operators defining the spherical Wigner function in Eq.~\eqref{WSphere} in Methods must be transformed accordingly via $(\,\cdot\,)\rightarrow \hat K(\,\cdot\,)\hat K^{\dagger}$.

It is now possible to define a generalized CV limit by introducing the unitarily modified number operator \( \hat{\tilde n}_a = \hat K\, \hat a^{\dagger}\hat a\, \hat K^{\dagger} \). The expectation value \( \langle \hat{\tilde n}_a \rangle{}_s= {}_s\!\bra{\psi}\hat{\tilde n}_a\ket{\psi}_s \) retains a clear interpretation as an effective excitation number relative to the chosen quantum reference frame. We then define the generalized CV subspace \( {\cal C} \) as the set of states \( \ket{\psi}_s \) such that \( \langle \hat{\tilde n}_a \rangle_s  \ll \sqrt{N} \). This condition selects states with small excitation number with respect to a new quantum reference, and the resulting subspace \( {\cal C} \) depends explicitly on the choice of \( \hat K \). In general, the induced space partition between physical states in this generalized CV limit and the remaining of the SSRC space that do not have a CV limit differ from the one associated with the original mode decomposition defined by \( \hat a \) and \( \hat b \) if \( \hat K \notin SU(2) \).

Within this framework, the generalized CV limit also defines a planar phase space corresponding to the tangent plane of the Bloch sphere generated by \( \{\hat{\mathcal J}_x,\hat{\mathcal J}_y,\hat{\mathcal J}_z\} \) at the point \( \hat K\ket{0}_a\ket{N}_b = \ket{0,N}_K \), which plays the role of the vacuum. Different choices of \( \hat K \) therefore correspond to different phase references and lead to distinct planar phase-space representations. In practice, these choices may be guided by experimental considerations—for instance, the quadrature phase space is naturally associated with the homodyne detection, but different physical regimes may be better represented in different planar spaces - or fundamental ones: varying \( \hat K \) allows one to probe the classical limit associated with different quantum reference frames \cite{CastroRuiz2025,Ballesteros2021,delaHamette2020,Giacomini2019}.

Finally, it is clear that the vicinity to the generalized tangent plane does not coincide, in general, with the one obtained for $\hat K=\mathbb 1$. The two CV limits describe equivalent physical subspaces only if $\langle\hat{\tilde n}_a\rangle_{K}\ll\sqrt{N}$ is equivalent to $\langle \hat n_a\rangle\ll\sqrt{N}$. In general, the states $\ket{n, N-k}_K$ are superpositions in the form \eqref{state}, involving both physical modes $a$ and $b$, and part of the information associated with operator $\hat {\tilde n}_b \equiv \hat K \hat b^{\dagger}\hat b \hat K^{\dagger}$, which acts on the quantum phase reference Hilbert space, again becomes implicit in the CV limit, through the approximations as $\hat {\tilde n}_b\ket{n, N-n}_{K}\approx N \ket{n, N-n}_{K}$ (see Methods).

Moving to a unitarily equivalent SSRC representation also modifies the definition of the planar Wigner function, and displacements in the planar tangent phase space are generated at the CV limit by the operators $e^{i\hat{\mathcal J}_x\theta}$ and $e^{i\hat{\mathcal J}_y\theta}$, with $\hat{\mathcal J}_x \rightarrow \sqrt{N}\,\hat K\hat x\hat K^{\dagger}$ and $\hat{\mathcal J}_y \rightarrow \sqrt{N}\,\hat K\hat p\hat K^{\dagger}$. 
The generalized planar phase space emerges from the spherical SSRC representation as the tangent plane at a point defining the vacuum, or equivalently the phase reference. While the mathematical procedure defining this limit and the associated phase-space distributions is independent of this tangent point, the states and operations phase-space properties, as negativities, depend explicitly on this choice of this reference, which is determined by the operator $\hat K$. In particular, $\ket{0,N}_K$, for instance, the new vacuum, is a Gaussian state in the generalized planar phase-space representation, irrespective of their form in the standard quadrature phase space (corresponding to $\hat K=\mathbb{1}$).
 This illustrates the relativity of the Gaussian/non-Gaussian dichotomy, that is also relative in the SSRC representation, where universal operations may be expressed in different unitarily equivalent ways for which the SG/SNG partition is not necessarily relevant. In addition, the planar generalized phase-space description exhibits, as the quadrature one, two key features. First, relevant physical properties—including phase-space negativities—may be progressively lost in the CV limit compared to the underlying SSRC representation \cite{moulonguet2026}. Second, as mentioned, this limit describes a constrained subspace of the full bosonic Hilbert space, in which phase-reference information is implicitly encoded. As a result, a direct identification between physical and computational resources within a purely CV framework, as commonly assumed in the literature \cite{davis2024, PhysRevResearch.5.L032006, PhysRevA.102.022411, PhysRevA.104.022408, PhysRevLett.125.040501, calcluth2023sufficient, ModularMenicucci, PhysRevA.104.012430, PRXQuantum.5.010331}, is generally incomplete, even for arbitrary $\hat K$. This observation clarifies one of the motivations and the relevance of the global framework developed here. Despite these limitations, planar phase-space descriptions remain central in experimental settings, particularly in quantum optics. 

We now connect these results with the fact that the qudit's generalized Pauli operators and the computational basis may also be defined up to a unitary transformation, as previously mentioned. Specifically, when the computational basis is defined as in Eq. \eqref{U}, $\ket{\overline j} \rightarrow \hat U\ket{j}_a\ket{N-j}_b $, one can set $\hat Z_U=\hat U\hat Z\hat U^{\dagger}$, $\hat F_U=\hat U\hat F\hat U^{\dagger}$, and $\hat X_U=\hat F_U\hat Z_U\hat F_U^{\dagger}$, where $\hat U$ is arbitrary. 
Of course, $\hat U$ could be in principle absorbed into $\hat K$ and vice-versa, since $\hat Z_{K U}=\hat K\hat U\hat Z \hat U^{\dagger} \hat K^{\dagger}=Z_{\tilde {\hat U}}$, with $\tilde {\hat U}=\hat K\hat U$. However, keeping them independent allows us to study how different encodings (choices of $\hat U$) appear in different CV physical spaces (determined by $\hat K$).

In order to establish the relation between physical and computational phase spaces, we define the SSRC unitary $\hat{\mathcal K} = \hat U^{\dagger} \hat K$. In particular, when $\hat{\mathcal K} = \mathbb{1}$, the correspondence between the qudit computational basis and the universal gate set coincides with that of the construction presented in Section~\ref{sec:Info}, which extends the Fock-state encoding of Ref.~\cite{PhysRevA.65.052316} to the SSRC framework (with $\hat U = \hat K = \mathbb{1}$). In this case, in computational basis states are represented by nonnegative distributions in the toric discrete computational phase space, while their physical counterparts correspond to Fock states, which exhibit negativities in the quadrature phase-space representation. This behavior persists for all choices of $\hat U$ and $\hat K$ such that $\hat{\mathcal K} = \mathbb{1}$. More generally, for any encoding ({\it i.e.}, any choice of $\hat U$), one can always select a planar phase-space representation ({\it i.e.}, a choice of $\hat K$) in which states and operations that are efficiently classically simulable—and therefore non-negative in the computational phase space—appear as negative distributions in the physical phase space.

This establishes point~\ref{two} and provides a systematic procedure to construct corresponding computational and physical phase-space representations for arbitrary encodings.

We now demonstrate point~\ref{one} by considering an explicit example with $\hat{\mathcal K} \neq \mathbb{1}$. We fix the planar phase space by choosing $\ket{0}_a\ket{N}_b$ as the reference point (vacuum) defining the tangent plane, as in Fig.~\ref{fig1}, and set $\hat K = \mathbb{1}$ and $\hat U = \hat R(\pi,0) = e^{i\hat J_y \pi}$. This choice yields $\hat Z_x =\hat U\hat Z  \hat U^{\dagger}  = e^{i\hat J_x \frac{2\pi}{N+1}}$, with eigenstates $\ket{j}_{a_x}\ket{N-j}_{b_x}$ forming the computational basis, i.e., $\ket{\overline{j}} \mapsto \ket{j}_{a_x}\ket{N-j}_{b_x}$. In the CV limit, these states map to quadrature eigenstates.

In this encoding, the operator $\hat X_x = \hat F_x \hat Z_x \hat F_x^{\dagger}$ acts as a shift, $\hat X_x \ket{n}_{a_x}\ket{N-n}_{b_x} = \ket{n+1}_{a_x}\ket{N-n-1}_{b_x}$, and can be written as $\hat X_x = e^{i\hat \theta_x \frac{2\pi}{N+1}}$. While $\hat X_x$ is a Pauli-like logical gate, it corresponds to a physically SNG operation. Nevertheless, when restricted to states in the CV limit—defined by $\langle \hat a^{\dagger}\hat a \rangle \ll N$ in the physical modes $(a,b)$—one finds $\hat X_x \approx e^{i\sqrt{\frac{2}{N}}\,\hat p}$, corresponding to an infinitesimal phase-space displacement.

This observation has two important consequences. First, within the CV subspace, logical operations that preserve positivity in the computational (discrete) phase space may correspond to Gaussian operations in the physical quadrature phase space, even when their exact physical implementation is non-Gaussian. Thus, the same phase-space displacement arises both as the CV limit of SG rotations and as the effective action of SNG logical gates. This illustrates how physical signatures such as non-Gaussianity or negativity can be progressively washed out in the CV limit, as previously mentioned and demonstrated in details in \cite{moulonguet2026}.

Second, this construction clarifies the relation between computational and physical phase spaces. Unlike previous approaches, which rely on a $d \to \infty$ limit, here the CV regime corresponds to a constrained subspace of an arbitrarily large but finite-dimensional Hilbert space of dimension $d = N+1$. In this sense, the CV limit does not enlarge the computational space but instead restricts the physically accessible region within it when $N$ is large.

Finally, the identification of $\hat X_x$ and $\hat Z_x$ with quadrature displacements ties the computational basis to the planar phase-space representation defined by the chosen vacuum. In this regime, Clifford operations map to Gaussian transformations and therefore correspond to non-negative distributions in the physical planar phase-space. Since this construction extends to arbitrary encodings (choices of $\hat U$) and choices of planar phase space (choices of $\hat K$) satisfying $\hat{\mathcal K} = \hat R(\pi,0)$, it establishes point~\ref{one}: for any encoding, one can construct a planar physical phase space in which all efficiently classically simulable (Clifford) operations are represented by non-negative distributions. Moreover, we have provided a systematic procedure to construct such phase-space representations for arbitrary encodings.

\section{Discussion}


We have introduced a general framework that unifies physical and computational phase-space representations of bosonic systems in arbitrary dimension, by explicitly promoting the phase reference to a quantum resource. A key consequence of this approach is that the full bosonic Hilbert space of a finite number of modes can be exactly mapped onto a quantum computational space, and therefore efficiently simulated by a quantum computer—an observation that is not apparent within the standard CV perspective, that we show to be a restriction to the physical and computational subspaces.

Within the SSRC representation, any bosonic basis naturally defines a qudit encoding of matching dimension, equipped with a discrete Wigner function whose computational properties—most notably, the efficient classical simulability of protocols—are independent of the underlying physical implementation. At the same time, we have shown that, for any encoding, one can construct planar physical phase-space representations in which these computational properties are faithfully reflected within the subspace defined by the CV limit.

The interplay between physical and computational descriptions is governed by the operator $\hat{\mathcal K} = \hat K \hat U$, where $\hat U$ specifies the encoding and $\hat K$ defines the phase reference (vacuum) of the physical phase space. This separation provides a unified way to compare different encodings and their physical realizations. Crucially, treating the phase reference as a quantum degree of freedom has no counterpart in conventional approaches, where it is typically assumed to be classical and the bosonic space is effectively truncated so as to be associated with a computational basis.

Altogether, our results formalize and clarify the roles of physical and computational resources in bosonic quantum information from a phase-space perspective, and provide a systematic framework to design and analyze bosonic quantum architectures, as well as the classical limits of bosonic quantum reference frames.


\section*{Acknowledgements}

We acknowledge funding from the Plan France 2030 through the project ANR-22-PETQ-0006 and from ANR-24-CE97-0003 EQUIPPS. We are also indebted to N. Fabre, J.-P. Gazeau, J. Lorgeré, N. Moulonguet, A. Leverrier, G. Ferrini, A. Maltesson, C. Calcluth, L. Rodung and R. G. Ahmed for inspiring discussions.

\section*{Methods}

\subsection*{Wigner functions: planar, spherical and discrete}

We define here the different Wigner functions used in this paper. 

The spherical Wigner function that contracts into the quadrature one in the CV limit \cite{PhysRevA.63.012102} is defined as:
\be\label{WSphere}
W_S(\Omega) = \Tr[\hat \rho_S \hat \Delta(\theta,\phi))],
\ee
where 
$\hat \rho_S=\sum_{n,n'}^N c_{n,n'}\ket{n}_a\ket{N-n}_b{}_b\bra{N-n'}{}_a\bra{n'}$ is the SSRC density matrix and $\hat \Delta(\theta,\phi)= e^{i\hat J_z\phi}e^{i\hat J_x\theta} \hat \Delta(0,0)e^{-i\hat J_z\phi}e^{-i\hat J_x\theta}$ and $\hat \Delta(0,0)=\sum_{l=0}^N\sum_{k=0}^N T_{l,k}^{(N)}$, with 
\be
T_{l,k}^{(N)}=\sqrt{\frac{l+1}{N+1}}\sum_{n,n'}^N\left< \begin{matrix} N~l \\ n~k \end{matrix}  \middle |\begin{matrix} N \\ n' \end{matrix} \right>\ket{n}_a\ket{N-n}_b{}_a\bra{n'}{}_b\bra{N-n'},
\ee
where $\left< \begin{matrix} N~l \\ n~k \end{matrix}  \middle |\begin{matrix} N \\ n' \end{matrix} \right>$ are the Clebsch-Gordan coefficients.

The planar Wigner function in the quadrature representation, can be obtained by defining $\hat P_0=\frac{1}{4\pi} \int d^2 \alpha \hat D(\alpha)$, where $\hat D(\alpha)=e^{\alpha \hat a^{\dagger}-\alpha^*\hat a}$ is the phase space displacement operator and $\hat P_0$ is the parity operator at the origin of the CV phase space. The Wigner function is then defined as 
\be\label{WPlanar}
W_P(\alpha)={\rm Tr}[\hat \rho_P \hat D(\alpha)\hat P_0 \hat D^{\dagger}(\alpha)],
\ee
where $\hat \rho_P=\sum_{n,n'}^{\infty}c_{n,n'}\ket{n}\bra{n'}$ is the density matrix in the CV representation. 

Finally, we define the discrete Wigner function of a qudit of dimension $d$ from the generalized Pauli operators $\hat T_{n,m} = \omega_d^{nm}\hat X^n \hat Z^m$, $n,m \in \mathbb{Z}_d\def \hat T_{\bf k}$, ${\bf k}=n,m$ as
\be\label{WQudit}
W({\bf k})= \frac{1}{d}{\rm Tr}[\hat \Delta_{\bf k} \hat \rho_D], 
\ee
with 
\be
\hat \Delta_{\bf k}=\hat T_{\bf k}\hat \Delta_0 \hat T_{\bf k}^{\dagger}, ~~~\hat \Delta_0= \frac{1}{d}\sum_{\bf k} \hat T_{\bf k},
\ee
where $\hat \rho_D = \sum_{j,j'=1}^d c_{j,j'}\ket{j}\bra{j'}$ is the density matrix in the qudit's space. 

\smallskip
\subsection*{The notation $\hat K\ket{n}_a\ket{N-n}_b=\ket{n, N-n}_K$: definition and examples}

The action of an arbitrary unitary transformation $\hat K$ on the basis of $\hat J_z$'s eigenstates defines a new basis, defined in the main text as $\hat K\ket{n}_a\ket{N-n}_b=\ket{n, N-n}_K$. Within this basis, it is still possible to identify independent subspaces, corresponding to partitions that are different from the one associated with $\hat a$ and $\hat b$ if $\hat K \notin SU(2)$. Such partitions can still be defined using the same restrictions as the ones identified as defining the CV limit. Defining ${\cal C}$ as the generalized CV limit subspace, we can define $\ket{\psi}_s~\in~{\cal C}$ iff $\langle \hat {\tilde n}_b \rangle_s \ll \sqrt{N}$, as in the main text. 

One example of a possible basis for which $\hat K \notin SU(2)$ is given by choosing $\hat K=\left ( e^{i\hat \theta_z\frac{2\pi}{(N+1)}}\right )^{N/2}$ ($N$ even), so that $\hat K\ket{0}_a\ket{N}_b=\ket{N/2}_a\ket{N/2}_b=\ket{0,N}_K$, that corresponds to an infinitely squeezed state in the CV limit, is the new vacuum.

\bibliography{bibPhaseSpace}

\end{document}